\documentclass[11pt,dvips]{article}
\usepackage{graphicx}
\usepackage{amssymb}
\usepackage{latexsym}

\begin{document}
\begin{titlepage}
\setcounter{page}{1}
\renewcommand{\thefootnote}{\fnsymbol{footnote}}

\begin{flushright}
ucd-tpg:09-05~~~~~\\
%arXiv:0911.xxxx
\end{flushright}

\vspace{5mm}
\begin{center}

{\Large \bf Electromagnetic Excitations of  $A_n$ Quantum Hall Droplets}
%: \\ Bulk and Edge Dynamics}

\vspace{0.5cm}

{\bf Mohammed Daoud}$^a${\footnote {\it Facult\'e des
Sciences, D\'epartement de Physique, Agadir, Morocco; email:
{\sf daoud@pks.mpg.de}}},  {\bf Ahmed Jellal}$^{b,c,d}$\footnote{{\sf jellal@pks.mpg.de} and
{\sf jellal@ucd.ma}}  and  {\bf Abdellah Oueld Guejdi}$^e$

\vspace{0.5cm}

$^a${\em  Max Planck Institute for Physics of Complex Systems,
N\"othnitzer Str. 38,\\
 D-01187 Dresden, Germany}
%\vspace{0.2cm}

{$^b$\em Physics Department, College of Sciences, King Faisal University,\\
PO Box 380, Alahsaa 31982,
Saudi Arabia}

{$^c$\em Saudi Center for Theoretical Physics, Dhahran,
Saudi Arabia}

$^d${\em  Theoretical Physics Group, Faculty of Sciences,
 Choua\"ib Doukkali University,\\
PO Box 20, 24000 El Jadida, Morocco} %\vspace{0.2cm}

$^e${\em Department of Mathematics, Faculty of Sciences, University Ibn Zohr,\\
PO Box 8106, Agadir,
Morocco}\\[1em]

\vspace{3cm}

\begin{abstract}

%We review the notion of $A_n$ statistics related to the class of simple Lie algebras $sl(n+1)$.

The classical description of $A_n$ internal degrees of freedom is given by making use of the
Fock--Bargmann analytical realization. The symplectic deformation of phase space,
including the internal degrees of freedom,
is discussed.  We show that the Moser's lemma provides a mapping to eliminate the fluctuations of
the symplectic structure, which become encoded in the Hamiltonian of the system. We discuss
the relation between Moser and Seiberg--Witten maps. One physics applications of this result
is the electromagnetic excitation of a large collection of particles, obeying the generalized
$A_n$ statistics, living in the complex projective space ${\bf CP}^k$ with $U(1)$
background magnetic field. We explicitly calculate the bulk and edge actions. Some particular
symplectic deformations are also considered.

\end{abstract}
\end{center}
\end{titlepage}

\newpage
%%%%%%%%%%%%%%%%%%%%%%%%%%%%%%%%%%%%%%%%%%%%%%%%
\section{Introduction}
%%%%%%%%%%%%%%%%%%%%%%%%%%%%%%%%%%%%%%%%%%%%%%%%
Quantum Hall effect in higher dimensions has  intensively been investigated in
the last decade from
different point of views [1-10]. This yielded interesting  results such as the
bosonization~[2, 11-13] that has been achieved by making use
of the incompressible Hall droplet picture.
In this framework,  the edge excitations of a quantum Hall droplet are described
by a generalized Wess--Zumino--Witten action.
Recently, the electromagnetic excitations of a quantum Hall droplet was discussed by
Karabali [11] and Nair [12] for the Landau systems in the complex projective space ${\bf CP}^k$.
In fact, the corresponding bulk and edge actions were derived [11]. Interestingly, it was shown that the bulk contribution
coincides with the $(2k+1)$-dimensional Chern--Simons action [12].

The main tool of these developments is the matrix formulation of the lowest Landau level Hall
droplet dynamics and the fact that the lowest Landau wavefunctions coincides with the $SU(k+1)$
coherent states. This provides a simple way to perform the semiclassical analysis in
the high magnetic field regime and spinless particles, which form a droplet
in the phase space. Along more or less similar lines, the inclusion of spin,
color or other degrees of freedom for particles in the phase space formulation
has been proposed by Polychronakos [13]. This leads to an elegant gauge generalization
of the usual semiclassical droplet picture for scalar particles. The Polychronakos
formalism gave rise also  the non linear higher dimensional  generalization of the
gauged Kac-Moody algebra.

Our aim is to analyze the quantum Hall effect in complex projective
spaces by incorporating the internal degrees of freedom in the phase space description.
We will particularly interested by the generalized $A_n$ statistics introduced firstly
by Palev [14], see also [15]. This kind of statistics is associated with
the classical Lie algebra $A_n$
and is similar from a mathematical point of view to the generalized spin systems introduced
by Randjbar-Daemi {\it et al.}  [16].
Our approach follows the same ideas developed in [13] but is some what different. Indeed,
we specify the nature of internal degrees of freedom and more importantly we realize
the coupling between the kinematical phase space and the internal degrees of freedom
via a symplectic modification procedure. This induces a noncommutativity between
two subspaces as well as  electromagnetic excitations of the Hall system.
We show that the symplectic fluctuations can be eliminated by dressing transformations
using the Moser lemma that is a refined version of the celebrated Darboux theorem.
In this way, the effect of the symplectic modification becomes encoded in the Hamiltonian
of the system.

The paper is organized as follows. In section 2, we recall the main tools needed
 for the forthcoming analysis. This can be done by defining
 the generalized statistics associated to the Lie algebras of class $A_n$
 and briefly  reviewing the quantum Hall effect
in ${\bf CP}^k$. In section 3, we give a complete description
of the analytical representation of $A_n$ statistics, which uses a representation \`a la
Fock--Bargmann and provides us with phase space description of the internal degrees of freedom
for particles obeying  $A_n$ statistics. In section 4, the symplectic structure of
the phase space of $A_n$ particles living in the manifold ${\bf CP}^k$ is introduced.
Using a matrix formulation, we derive the action governing the edge excitations of
the quantum Hall droplet in the whole phase space. We show that the inclusion of the internal
degrees of freedom do not gives rise to any nontrivial dynamics. Thus, in section 5,
we purpose a symplectic modification (deformation) of the phase space structure. Using
the Moser's lemma, we give explicitly the dressing transformation, which allows to deal
with un-deformed symplectic structure and include the electromagnetic coupling effects
in the excitation potential. We also discuss the relation between the Moser's lemma [17]
and  Seiberg-Witten map [18], see also [19-20]. In section 6, we give the bulk and
edge actions describing the $A_n$ particles in the presence of the electromagnetic
interactions. In section 7, we treat the case of particles with $A_1$ statistics
living in two-sphere. We consider the most simplest form of
the electromagnetic tensor field (matrix elements constants). In this particular case,
a dressing transformation based on the Hilbert--Schmidt orthonormalisation procedure
is used to eliminate fluctuations. Obviously, this agrees with one derived
from Moser's lemma for small fluctuations. We obtain the action describing the edge excitations
and show that, in this special situation, the velocities of the propagation of
edge field along the droplet becomes modified due to the electromagnetic coupling.
Concluding remarks and comments close this paper.

%%%%%%%%%%%%%%%%%%%%%%%%%%%%%%%%%%%%%%%%%%%%%%%%%%%%%%%%%%%%%%%%%%%%%%%%%%%%%%%%%%%%%%%%%%
\section{Specifics of $A_n$ statistics and quantum Hall effect in ${\bf CP}^k$}
%%%%%%%%%%%%%%%%%%%%%%%%%%%%%%%%%%%%%%%%%%%%%%%%%%%%%%%%%%%%%%%%%%%%%%%%%%%%%%%%%%%%%%%%%%

The notion of Bose and Fermi operators has been generalized many years ago
to parabosons and parafermions [21].  In this generalization of conventional
Bose and Fermi statistics, the paraboson or
parafermion algebra is generated by $n$ pairs of creation and annihilation operators $(a_{+i},
 , a_{-i} ), {\rm with} ~  i =1, 2, \cdots , n$. They satisfy the trilinear relations
which replace the standard bilinear commutation or
anti-commutation relations. These are
\begin{eqnarray}
&& \left[[a_{+i} , a_{-j}]_{\pm}, a_{-k} \right] = - 2\delta_{ik} a_{-j}\nonumber\\
&& \left[[a_{+i} , a_{+j}]_{\pm}, a_{-k} \right] = - 2\delta_{ik} a_{+j}\\
 && \left[[a_{-i} , a_{-j}]_{\pm}, a_{-k} \right] = 0\nonumber
\end{eqnarray}
where as usual $[x, y]_{\pm} = xy \pm yx $ and the sign $\pm$ refer, respectively,
to parabosons and parafermions.
The triple relations of parafermion operators give one possible realization of the orthogonal Lie
algebra $so(2n + 1) = B_n $ [22]. The para-Bose statistics are connected to the orthosymplectic
superalgebra $osp(1/2n) = B(0, n)$ [23]. In view of this connection between Lie algebras
and parastatistics, a classification of generalized quantum statistics were derived for
the classical Lie algebras  of class $ A_n, B_n, C_n $ and $D_n$  as well as
the classical Lie superalgebras [14-15, 24].

The $A_n$ statistics are defined
through of $n$ creation and annihilation operators $s_{\pm i}$ %($ i = 1, 2, \cdots, n$)
satisfying
\begin{eqnarray}
&&
\left[ [s_{+i} , s_{-j}] ,  s_{+k} \right] = +  \delta_{jk}
 s_{+i} +  \delta_{ij} s_{+k}
 \nonumber \\
&& \left[ [ s_{+i} , s_{-j}] , s_{-k} \right] =  -  \delta_{ik}
 s_{-j} -   \delta_{ij}  s_{-k}
\end{eqnarray}
implemented by the mutual commutation relations % between creation and annihilation operators, such as
\begin{equation}
[s_{+i} , s_{+j}] = 0,
\qquad
[ s_{-i} , s_{-j}]  =  0.
\end{equation}
It is interesting to stress that the formalism of Lie algebraic statistics is deeply related to
the notion of generalized spin systems introduced in [16] for an arbitrary compact group $G$.
In this paper, we consider particles obeying the $A_n$ statistics. The associated algebraic
structures will be given in the next section. In particular, we derive the analytical
realization of the corresponding representation space, which defines the internal phase space.
These particles are assumed to live in the $2k$-dimensional complex projective space
 ${\bf CP}^k$.

 For ${\bf CP}^k$ the magnetic field, which leads to the Landau levels, is proportional to
the symplectic two-form, which is
\begin{equation}
\omega_0 ( z , \bar z) = i  g_{i\bar j} dz^i\wedge d\bar z^j
\end{equation}
where the metric elements are
\begin{equation}
g_{i\bar j} = q (1 + \bar z\cdot z)^{-2}[(1 + \bar z\cdot z)\delta_{ij} - \bar z_iz_j]
\end{equation}
for $i,j = 0, 1, \cdots, k$. The lowest Landau levels (LLL) wavefunctions   [2]
\begin{equation}
| z_1, z_2, \cdots , z_k \rangle =  (1 + \bar z\cdot z)^{-\frac{q}{2}}
\sum_{m_1}\sum_{m_2}\cdots
\sum_{m_k} \sqrt{\frac{q!}{(q-m)!}}
\frac{z_1^{m_1}}{\sqrt{m_1!}}\frac{z_2^{m_2}}{\sqrt{m_2!}}\cdots \frac{z_k^{m_k}}{\sqrt{m_k!}}| m_1, m_2, \cdots , m_k \rangle.
\end{equation}
coincide  with the ${\bf CP}^k$ coherent states, with $ m = m_1+m_2+\cdots+m_k$. Since, the LLL
 constitute an over complete set with respect to the measure
 \begin{equation}
 d\mu ( \bar z , z) = \frac{(q+k)!}{\pi^k q!} ~ \frac{d^2z_1d^2z_2 \cdots d^2z_k}{(1+\bar z\cdot z)^{k+1}}.
\end{equation}
This provides us with an elegant way to perform the semiclassical analysis, which can be
done  by associating to any operator acting on the states space
 \begin{equation}
{\cal F}^{\rm LLL} = \{ \vert m_1, m_2, \cdots , m_k\rangle ~ ;~ m_1+m_2+\cdots+m_n \leq q \}
\end{equation}
 a function (or symbol) defined on the phase space coordinate. The commutator between any two operators gives the
 so-called Moyal star product. It coincides (up to multiplicative factor) with the Poisson bracket in the high
 magnetic field regime.

%%%%%%%%%%%%%%%%%%%%%%%%%%%%%%%%%%%%%%%%%%%%%%%%%%%%%%%%%%%%%%%%%%%%%%%%%%%%%%%%%%%%%%%%%%%%%%%%
\section{Generalized $A_n$ spin variables}
%%%%%%%%%%%%%%%%%%%%%%%%%%%%%%%%%%%%%%%%%%%%%%%%%%%%%%%%%%%%%%%%%%%%%%%%%%%%%%%%%%%%%%%%%%%%%%%%

The semiclassical description of internal degrees of freedom  (spin, color, flavor, etc, $\cdots$)
presents some conceptual problems because they not possess classical counter parts. However, there
exists many ways to realize the spin variables. For instance, the
 $SU(2)$ spin variables can be viewed as arising from  quantization of the coset space
 $SU(2)/U(1)$ with an appropriate canonical form. This can be generalized to
 the coset space ${\bf CP}^n = SU(n+1)/SU(n)\times U(1)$ and visualized as a realization of
 the LLL of a particle living in the group manifold
$SU(n+1)$ with a magnetic field proportional to the ${\bf CP}^n$ symplectic form.
Such realization %of generalized spin variables
was introduced in [16] where a
scheme to generalize the notion
of $SU(2)$ spin systems to an arbitrary compact group $G$ was given.
This generalization is based on the Holstein--Primakoff representation of spin matrices and used
 the coherent state method.

 In other words, the finite dimensional representation
spaces, on which the spin operators (generators of the group $G$) act, are provided with
an over-complete basis labeled by coordinates on a coset space $G/H$ where $H$ includes
the Cartan subgroup. According to [13, 16], it follows that the classical phase
space, encoding the internal quantum numbers of particles, can be realized by considering
the internal quantum numbers arising from the quantization of an internal compact phase
space for  particles. In this sense, we will show that the generalized $A_n$ statistics
algebra introduced by Palev [14], deeply linked to the notion of triple Lie systems introduced by Jacobson [25],
provides us with a nice way to define the generalized spin variables and
establish a connection between $A_n$ statistics and $SU(n+1)$ spin matrices.

%%%%%%%%%%%%%%%%%%%%%%%%%%%%%%%%%%%%%%%%%%%%%%%%%%%%%%%%%%%%%%%%%%%%%%%%%%%%%%%%%%%%%%%%%%%%%%%%
\subsection{$A_n$ spin systems}
%%%%%%%%%%%%%%%%%%%%%%%%%%%%%%%%%%%%%%%%%%%%%%%%%%%%%%%%%%%%%%%%%%%%%%%%%%%%%%%%%%%%%%%%%%%%%%%%

We start by introducing
the $A_n$ spin systems. Indeed, they can be defined
in terms of the generators $s_{\pm 1}, s_{\pm 2}, \cdots , s_{\pm n}$
verifying the triple relations
\begin{eqnarray}
&&
[ s_{+i} , s_{-j}  , s_{+k} ] = +  \delta_{jk}
 s_{+i} +  \delta_{ij} s_{+k}
\\
&& [ s_{+i} , s_{-j}  , s_{-k} ] =  -  \delta_{ik}
 s_{-j} -   \delta_{ij}  s_{-k}
\\
&& [ s_{+i} , s_{+j}  , s_{\pm k} ] =  [ s_{-i} , s_{-j}  , s_{\pm k} ] = 0.
\end{eqnarray}
This description of $A_n$ Lie algebras  is a particular
case of describing Lie algebras through triple systems initiated by Jacobson [25].  Recall
that, a Lie triple system is defined as a subspace of an associative algebra that is closed
under the ternary composition $[ x , y , z]$. They  satisfy the conditions
\begin{eqnarray}
&& [ x , y , z] = - [ y , x, z]
\\
&& [ x , y , z] + [ y , z , x] + [ z , x , y] = 0
\\
&& [ u, v, [ x , y , z]] = [[u, v, x] , y , z] + [ x, [u , v , y] , z]+[ x , y, [ u, v  , z]].
\end{eqnarray}
It is easy to see  that the ternary composition
\begin{equation}
[ x , y , z] =  (xyz) - (yxz), \qquad  \qquad (xyz) = xyz + zyx
\end{equation}
 verifies the conditions (12-14). In order to check that
 the generators $s_{\pm i}$ close a Lie triple algebra under the composition operation (15),
 it is convenient to consider $A_n$ as subalgebra of the Lie algebra $gl(n+1)$
 spanned by the generators $e_{ij}$, with $e_{ij}e_{kl} = \delta_{ik} e_{il}$.
 This can be used to realize  $s_{\pm i}$ as
\begin{equation}
 s_{+i} = e_{i0}\qquad s_{-i} = e_{0i}.
\end{equation}
Thus, we obtain
\begin{equation}
(s_{+i}s_{-j}s_{+k}) = \delta_{ij}s_{+k} + \delta_{kj} s_{+i}, \qquad (s_{-i}s_{+j}s_{-k}) = \delta_{jk}s_{-i} + \delta_{ij} s_{-k}
\end{equation}
with all other triples are vanishing, now it is clear that (9-10)
are verified. Obviously for
$[ x , y] = x y - y x$, any Lie algebra is a Lie triple system. Indeed,  (15) can be rewritten
as
$$ [ x , y , z] = [[x , y] , z ]$$
which coincides with the definition of generalized $A_n$ statistics (2-3).
From a purely algebraic point of view, the Jacobson formulation provides an
alternative way to the Chevalley description of classical Lie algebras.
From a physical point of view, the Jacobson generators can be seen as creation and annihilation
operators satisfying triple relations.

We now consider an Hilbertean representation of the algebra $A_n$. Let  ${\cal F}^{\rm spin}$
be the Hilbert--Fock space on which the
generators act. Since the algebra  is
spanned by $n$ pairs of Jacobson generators, it is natural to
assume that the Fock space is given by
\begin{equation}
{\cal F}^{\rm spin} = \{ |l_1, l_2,\cdots,\ \  l_n\rangle\ , l_i \in {\mathbb N}\}
\end{equation}
The operators $s_{\pm i}$ are acting on  ${\cal F}^{\rm spin}$ [14], see also [26], as
\begin{equation}
s_{-i} |l_1,\cdots, l_i,\cdots , l_n\rangle\ = \sqrt{l_i (p+1 -(l_1+l_2+\cdots+l_n))}|l_1,\cdots, l_i-1,\cdots , l_n\rangle
\end{equation}
\begin{equation}
s_{+i} |l_1,\cdots, l_i,\cdots , l_n\rangle\ = \sqrt{(l_i+1) (p
-(l_1+l_2+\cdots+l_n))}|l_1,\cdots, l_i+1,\cdots , l_n\rangle.
\end{equation}
The dimension of the irreducible representation space
is determined by the  condition
\begin{equation}
p+1 -(l_1+l_2+\cdots + l_n) > 0.
\end{equation}
${\cal F}^{\rm spin}$ is generated by  a finite number of states satisfying the
condition $l_1+l_2+\cdots+l_n \leq p$, which is a generalized version of
the exclusion Pauli principle
 [14]. This means that no more than $p$ particles can be accommodated in the same quantum state.
 The integer $p$ defines the order of the statistics. The Fock space dimension is given by
$$ {\rm dim} {\cal F}^{\rm spin} = \frac{(p+n)!}{p!n!}.$$

Finally, we point out one
interesting property of the  $A_n$ systems.
Indeed, let us introduce the operators
$$b_i^{\pm} = \frac{s_{\pm i}}{\sqrt{p}}$$
where $i = 1, 2, \cdots, n$ and consider $p$ very large. From
(19) and (20), one can check that  the $A_n$ algebra coincides with  $n$-copies
of the Weyl--Heisenberg algebras. In addition, the Jacobson or spin
operators reduce to the standard bosonic creation and annihilation operators.
Note also that, for $n = 1$ and $p=1$, one recovers the Fermi creation and annihilation operators.

%%%%%%%%%%%%%%%%%%%%%%%%%%%%%%%%%%%%%%%%%%%%%%%%%%%%%%%%%%%%%%%%%%%%%%%%%%%%%%%%%%%%%%%%%%%%%%%%
\subsection{Geometry of $A_n$ spin systems}
%%%%%%%%%%%%%%%%%%%%%%%%%%%%%%%%%%%%%%%%%%%%%%%%%%%%%%%%%%%%%%%%%%%%%%%%%%%%%%%%%%%%%%%%%%%%%%%%

To define the spin variables of generalized $A_n$ systems, we use the coherent state method
developed in [16], see also [26-27]. In fact, we will outline the main features needed for our task.
The required realization uses a suitably
defined Hilbert space of entire analytical functions.

The Jacobson
annihilation generators $s_{- i}$ are realized as first order
differential operators with respect to a complex variables $w_i$. This is
\begin{equation}
s_{-i} \longrightarrow \frac{\partial}{\partial w_i}.
\end{equation}
The key point of such analytical realization lies on the fact that we represent the Fock
 states $\vert l_1, l_2, \cdots, l_n \rangle$
as power of complex variables $w_1, w_2, \cdots, w_n$, such as
\begin{equation}
| l_1, l_2 \cdots , l_n\rangle  \longrightarrow C_{ l_1,\cdots
,l_n} w_1^{l_1} w_2^{l_2} \cdots w_n^{l_n}.
\end{equation}
Using  the annihilation operator actions on
${\cal F}^{\rm spin}$ and  correspondences (22-23), the coefficients $C_{ l_1,l_2,
\cdots,l_n}$ are obtained (up to a normalization constant) as
\begin{equation}
C_{ l_1,l_2,
\cdots,l_n}= \sqrt{{\frac{p!
}{(p-l)!}}}\frac{1}{\sqrt{l_1!}\sqrt{l_2!}\cdots \sqrt{l_n!}}
\end{equation}
where $l = l_1 + l_2 + \cdots + l_n$. Using
(20) and (24), one can determine the differential action of the
Jacobson creation operators, which is
\begin{equation}
s_{+i} \longrightarrow p w_i - w_i \sum_{j=1}^n w_j\frac{d}{dw_j}.
\end{equation}
Note that, the Jacobson generators act as first order linear differential
operators.

On the other hand, an  arbitrary vector
$|\psi \rangle =
\sum_{l_1}\sum_{l_2}\cdots
\sum_{l_n}\psi_{l_1, l_2 \cdots , l_n}| l_1,l_2,
\cdots ,l_n \rangle$ of ${\cal F}^{\rm spin}$
 is realized as
\begin{equation}
\psi(w_1, w_2 \cdots , w_n) =
\sum_{l_1}\sum_{l_2}\cdots
\sum_{l_r} \psi_{l_1, l_2 \cdots , l_n} C_{l_1, l_2
\cdots , l_n} w_1^{l_1} w_2^{l_2} \cdots w_n^{l_n}.
\end{equation}
%One can write the function $\psi(w_1, w_2 \cdots , w_n)$
It can be written as the product of the state
$|\psi\rangle $ with some ket $\langle \bar w \vert := \langle  \bar w_1, \bar w_2 \cdots ,
\bar w_n \vert$ labeled by the complex conjugate of the
variables $w_1, w_2, \cdots , w_n$. This is
\begin{equation}
\psi(w_1, w_2, \cdots , w_n)= {\cal N} \langle  \bar w_1, \bar w_2,
\cdots , \bar w_n |\psi \rangle
\end{equation}
where ${\cal N}$ is the normalization constant. Taking $|\psi\rangle = | l_1, l_2, \cdots , l_n \rangle$, we
have
\begin{equation}
\langle  \bar w_1, \bar w_2,
\cdots , \bar w_n\vert l_1, l_2, \cdots, l_n\rangle = {\cal N}^{-1} C_{l_1, l_2, \cdots, l_n}w_1^{l_1} w_2^{l_2} \cdots w_n^{l_n}.
\end{equation}
This implies
\begin{equation}
| w_1, w_2, \cdots , w_n \rangle =  {\cal N}^{-1}
\sum_{l_1}\sum_{l_2}\cdots
\sum_{l_n} \sqrt{\frac{p!}{(p-l)!}}
\frac{w_1^{l_1}}{\sqrt{l_1!}}\frac{w_2^{l_2}}{\sqrt{l_2!}}\cdots \frac{w_n^{l_n}}{\sqrt{l_n!}}| l_1, l_2, \cdots , l_n \rangle.
\end{equation}
The normalization constant  is given by
\begin{equation}
{\cal N} = \left(1 +|w_1|^2+ |w_2|^2+ \cdots + |w_n|^2\right)^{\frac{p}{2}}
= \left(1 + \bar w \cdot w\right)^{\frac{p}{2}}.
\end{equation}

The states (29) are continuous in the labeling, constitute an over
complete set with respect to the measure, which can be evaluated as
 \begin{equation}
 d\mu ( \bar w , w) = \frac{(p+n)!}{\pi^n p!} ~ \frac{d^2w_1d^2w_2 \cdots d^2w_n}{(1+\bar w\cdot w)^{n+1}}
\end{equation}
and then
give the $SU(n+1)$ coherent states. This analytical realization enables us to define the spin variables
of the $A_n$ phase space systems. It is equipped with
a symplectic (Kahler) two-form that makes it into classical phase space.
This is realized by introducing the Kahler
potential
 \begin{equation}
K_0( \bar w , w) ~ =
~ {\rm Ln} \vert \langle 0 \vert w \rangle \vert^{-2}  = p {\rm Ln}( 1 + \bar w\cdot w)
\end{equation}
to define the required closed symplectic two-form as
\begin{equation}
 \omega_0 (w , \bar w)= i g_{i\bar j} dw^i\wedge d\bar w^j.
\end{equation}
The corresponding Poisson bracket is
\begin{eqnarray}
\{ f , g \} = -i g^{i\bar j} \bigg(\frac{\partial f}{\partial
w_i}\frac{\partial g}{\partial \bar w_j} - \frac{\partial
g}{\partial w_i}\frac{\partial f}{\partial \bar w_j}\bigg).
\end{eqnarray}
The components of the metric tensor are
$$ g_{i \bar j} = \frac{\partial^2 K_0(\bar w , w )}{\partial w_i \partial \bar w_j} =
p (1 + \bar w\cdot w)^{-2}[(1 + \bar w\cdot w)\delta_{ij} - \bar w_iw_j]$$
and  the matrix elements of its inverse are given by
$$ g^{i\bar j} = \frac{1}{p} (1 + \bar w\cdot w) (\delta_{ij} + w_i\bar w_j).$$
As excepted, this is precisely the Kahler structure of the complex projective spaces ${\bf CP}^n$.

%%%%%%%%%%%%%%%%%%%%%%%%%%%%%%%%%%%%%%%%%%%%%%%%%%%%%%%%%%%%%%%%%%%%%%%%%%%%%%%%%%%%%%%%%%%%%%%%
\section{ Hall droplet with internal degrees of freedom}
%%%%%%%%%%%%%%%%%%%%%%%%%%%%%%%%%%%%%%%%%%%%%%%%%%%%%%%%%%%%%%%%%%%%%%%%%%%%%%%%%%%%%%%%%%%%%%%%

%%%%%%%%%%%%%%%%%%%%%%%%%%%%%%%%%%%%%%%%%%%%%%%%%%%%%%%%%%%%%%%%%%%%%%%%%%%%%%%%%%%%%%%%%%%%%%%%
\subsection{ Phase space description}
%%%%%%%%%%%%%%%%%%%%%%%%%%%%%%%%%%%%%%%%%%%%%%%%%%%%%%%%%%%%%%%%%%%%%%%%%%%%%%%%%%%%%%%%%%%%%%%%

Most of discussions about
%It is common in discussions of
the quantum Hall effect  ignore the fermionic internal degrees
of freedom. Recently, an elegant way to take into account of them
 was proposed by Polychronakos [13]. According to his ideas,
 we will consider
 a dense collection of noninteracting particles living in ${\bf CP}^k$.
 This defines the kinematical phase space  that we enlarge  to include the $A_n$ spin variables.
The total phase space is then a direct product of the usual phase space and the additional one encoding
the internal degrees of freedom of the particles. This is
$${\cal M} = {\bf CP}^k \times {\bf CP}^n$$
where its   dimension is $2(k+n)$. It is endowed with a symplectic two-form of type
\begin{equation}
\omega_0= i g_{i\bar j}(\bar w, w) dw^i\wedge d\bar w^j  + i g_{i\bar j}(\bar z, z) dz^i\wedge d\bar z^j.
\end{equation}
Clearly, our choice of $A_n$ spin systems living in ${\bf CP}^k$ will help us
to do our task because the internal and kinematical phase spaces have identical
geometrical structures. Moreover, our analysis can also be applied to other
systems, for instance, like particles obeying $B_n$, $C_n$ or $D_n$ statistics
living
in the Grassmanian manifold $ U(k)/U(k_1) \times U(k_2)\times \cdots \times U(k_t)$,
with  $k_1 + k_2 +\cdots + k_t = k$.

To determine the dynamics of the system under consideration, for a large magnetic field ($ q $ large),
some tools are needed. These concern the notion of star product, Hushimi (or density)
distribution and classical Hamiltonian. We assume also that the order $p$ of the $A_n$ statistics is large.
The first needed ingredient is the star product between two functions, which is defined as
the mean value of the product of two operators. More precisely, one associate to every operator $A$
acting on the Fock space ${\cal F}^{\rm LLL} \otimes{\cal F}^{\rm spin}$, the function
\begin{equation}
{\cal A}(\bar z, \bar w, z , w) = \langle z ,  w | A | z , w \rangle.
\end{equation}
An associative star product of two functions ${\cal A}$
and ${\cal B}$ is defined by
\begin{equation}
{\cal A}\star {\cal B} = \langle z , w | AB | z , w
\rangle = \int d\mu(\bar{z'}, z') d\mu(\bar{w'}, w') \langle z , w | A | z' ,  w' \rangle\langle z' , w'| B | z , w\rangle
\end{equation}
where the measures are given by (7) and (31). It follows that the star product between two functions
on the  phase space  can be evaluated to get [27]
\begin{equation}
{\cal A}\star {\cal B} = {\cal A}
{\cal B} - g^{i \bar j}
\frac{\partial{\cal A}}{\partial{w_i}}\frac{\partial{\cal B}}{\partial{\bar w_j}} - g^{i \bar j}
\frac{\partial{\cal A}}{\partial{z_i}}\frac{\partial{\cal B}}{\partial{\bar z_j}}.
\end{equation}
Then, the symbol or function associated with the commutator of two
operators $A$ and $B$ is given by
\begin{equation}
\langle z ,  w |[ A , B] | z , w\rangle = \{{\cal A}, {\cal B}\}_{\star} =  - g^{i \bar j}\bigg(
\frac{\partial{\cal A}}{\partial{w_i}}\frac{\partial{\cal B}}{\partial{\bar w_j}}
- \frac{\partial{\cal B}}{\partial{w_i}}\frac{\partial{\cal A}}{\partial{\bar w_j}}\bigg) - g^{i \bar j}\bigg(
\frac{\partial{\cal A}}{\partial{z_i}}\frac{\partial{\cal B}}{\partial{\bar z_j}}
- \frac{\partial{\cal B}}{\partial{z_i}}\frac{\partial{\cal A}}{\partial{\bar z_j}}\bigg)
\end{equation}
where
\begin{equation}
 \{{\cal A} , {\cal B}\}_{\star} = {\cal A}\star {\cal B} -
{\cal B} \star {\cal A}
\end{equation}
is the so-called the Moyal bracket.

It is well understood that, for large magnetic field, a large collection of particles
in the LLL (partially filled) behaves like an incompressible droplet.
In the present context, each available state is also labeled by the $A_n$ spin quantum numbers.
Similarly, to the spinless case, we define the density operator as
 \begin{equation}
\rho_0 = \rho_0^{\rm LLL} \otimes \rho_0^{\rm spin}
\end{equation}
where $\rho_0^{\rm LLL}$ is the density matrix of $M$ particles in the LLL
and the matrix $\rho_0^{\rm spin}$ incorporates the information about the $A_n$ spin states.
The  symbol associated with the density matrix  can be expressed in the coherent state basis
to verify that it is a function of $\bar z \cdot z$ and $\bar w \cdot w$ and with a spherical
finite spatial extent. This tells us that we should parameterize the boundary of the droplet
by a function $ r = r (q ~ \bar z\cdot z ,  p ~ \bar w\cdot w)$, such as
  \begin{equation}
\rho_0 (\bar z\cdot z, \bar w\cdot w) = \Theta (M - r^2)
\end{equation}
where $\Theta$ is the usual step function. We have a spherical droplet with a radius proportional to $\sqrt{M}$.

The edge dynamics of spinless droplet is described by the WZW  action of type in higher dimensional spaces [2].
According to these previous works, the dynamics of the droplet (42) can be characterized in the following way.
One starts with the diagonal density matrix $\rho_0$ with $M$ states occupied
$(M < {\rm dim}{\cal F}^{\rm LLL}\times {\rm dim}{\cal F}^{\rm spin} )$,
then the fluctuations preserving the number of states are given by unitary transformation
\begin{equation}
\rho_0 \longrightarrow \rho = U \rho_0 U^{\dagger}
\end{equation}
and the equation of motion is the quantum Liouville equation
\begin{equation}
i \frac{\partial \rho}{\partial t} = [ V , \rho].
\end{equation}
The operator $U$ contains
information concerning the edge dynamics  and the operator $V$ is the potential generating
the excitations on the boundary of the droplet. We consider an oscillator like confining
potential as in the spinless case. This  is given by
\begin{equation}
 {\cal V}(\bar z , \bar w, z , w ) = \langle z , w |V| z , w \rangle = q ~ \frac{\bar z
 \cdot z}{1+\bar z \cdot z} + p ~ \frac{\bar w
 \cdot w}{1+\bar w \cdot w} .
\end{equation}
Here, we have restricted our choice of the quadratic excitation in terms of the internal
space variables. This is due to the fact that we choose $V$ so that $[ V , \rho_0 ] = 0$.
For large $q$ and $p$, the excitation potential (45) goes to a superposition of harmonic oscillators.
Note that, if we consider an equilibrium configuration, the droplet fills up the phase space
to an energy level $V_0$. That is the boundary $r$ is such that ${\cal V}(r)$ is constant.
This defines the droplet boundary as
\begin{equation}
r^2 = q ~ \bar z\cdot z + p ~ \bar w\cdot w.
\end{equation}

%%%%%%%%%%%%%%%%%%%%%%%%%%%%%%%%%%%%%
\subsection{Edge effective action}
%%%%%%%%%%%%%%%%%%%%%%%%%%%%%%%%%%%%%%%%

The action that gives the dynamics of edge excitations is
\begin{equation}
S_0 = \int dt {\rm Tr} \left\{ \rho_0 U^{\dag}\left(i\partial_t - V\right)U \right\}.
\end{equation}
As expected, one can check that the minimization condition for this action leads to the
quantum Liouville equation (44). Along the method used by Sakita [28], see also [29],
it is more adequate to write the unitary operator $U$
\begin{equation}
U= e^{+i\Phi}, \qquad \Phi^{\dag} = \Phi
\end{equation}
in terms of the abelian field $\Phi$, which will determine the nature of the droplet
fluctuations. Reporting (48) in the above action, one can see, after some algebra, that
the kinetic part is given by
\begin{equation}
 i \int dt\ {\rm Tr}\left(\rho_0
U^{\dag}\partial_tU\right) \simeq - \frac {i}{2} \int dt d\mu (\bar z , z)\ d\mu (\bar w , w)\
\{\Phi,\rho_0\}_{\star}\partial_t\Phi
\end{equation}
where we have dropped the terms in $\frac{1}{q^2}$ and $\frac{1}{p^2}$ as well as that related to
the total time derivative. In (49), $\Phi = \Phi (\bar z, \bar w , z , w)$ stands for the symbol
associated with the operator $\Phi$. The potential energy can be expanded as
\begin{equation}
{\rm Tr}\left(\rho_0 U^{\dag} V U\right) = {\rm Tr}\left(\rho_0  V \right) + i {\rm Tr}\left([\rho_0, V] \Phi\right)
+ \frac{1}{2}{\rm Tr}\left([\rho_0, \Phi ][V, \Phi ]\right).
\end{equation}
The first term in r.h.s of (50) is $\Phi$-independent. We drop
it since does not contains any information about the dynamics of the
edge excitations. The second term vanishes, namely $[\rho_0 , V] = 0$. The last term can be
evaluated using the correspondence between commutators and Moyal brackets. Thus,
the action (47) rewrites as
\begin{equation}
S_0 \simeq -\frac {1}{2} \int dt d\mu (\bar z , z)\ d\mu (\bar w , w)\
\{\Phi,\rho_0\}_{\star}\left[ i\partial_t\Phi ~ - ~\{ {\cal V} , \Phi \}_{\star} \right].
\end{equation}
The Moyal brackets involved in the above action can easily be evaluated by making use
of (39), (42) and (45). % given the density function and the excitation potential.
Consequently, we obtain
\begin{equation}
S_0 \approx \frac{1}{2}\int  dt \ \left(\frac{\partial\rho_0}{\partial r^2}\right)
\left\{({\cal L}\Phi)( \partial_t\Phi )+ ({\cal
L}\Phi)^2\right\}
\end{equation}
where the first order differential operator $ {\cal L}$ is defined by
\begin{equation}
 {\cal L} =  i (1 - \bar w \cdot w)^2 \left(w \cdot \frac{\partial}{\partial
 w}
- \bar w \cdot \frac{\partial}{\partial \bar w}\right) + i (1 - \bar z \cdot z)^2
\left(z \cdot \frac{\partial}{\partial z} - \bar z \cdot \frac{\partial}{\partial \bar z}\right).
\end{equation}

Since the derivative of the density function is a delta function, the action is defined on
the boundary of the droplet and involved only the time derivative of $\Phi$ and the
tangential derivative ${\cal L}\Phi$. It
is similar to that derived in the spinless case  for
${\bf CP}^k$ [2], the Bergman ball ${\bf B}^k$  and flag manifold ${\bf F}_2$ [10].
More interestingly, the equation of motion arising from (52) for the edge field $\Phi$ is
\begin{equation}
 {\cal L} ( \partial_t \Phi + {\cal L} \Phi) = 0.
\end{equation}
This  shows
 that any nontrivial dynamics arose out in this description. The kinematical
 phase space and the internal spin degrees
of freedom dynamics are disconnected. In this respect, it is interesting to couple
spin and kinematical degrees of freedom. This coupling can be realized by introducing
a nonzero phase space structure between two relevant spaces. More precisely, this can be achieved
by a symplectic deformation of  two-from (35). %This is the main issue of the following section.

%%%%%%%%%%%%%%%%%%%%%%%%%%%%%%%%%%%%%%%%%%%%%%%%%%%%%%%%%%%%%%%%%%%%%%%%%%%%%%%%%%%%%%%%%%%%%%%%
\section{Noncommutative phase space and Seiberg--Witten map}
%%%%%%%%%%%%%%%%%%%%%%%%%%%%%%%%%%%%%%%%%%%%%%%%%%%%%%%%%%%%%%%%%%%%%%%%%%%%%%%%%%%%%%%%%%%%%%%%

%%%%%%%%%%%%%%%%%%%%%%%%%%%%%%%%%%%%%%%%%%%%%%%%%%%%%%%%%%%%%%%%%%%%%%%%%%%%%%%%%
\subsection{{Deformed symplectic structure}}
%%%%%%%%%%%%%%%%%%%%%%%%%%%%%%%%%%%%%%%%%%%%%%%%%%%%%%%%%%%%%%%%%%%%%%%%%%%%%%%%%

According to the total phase space realization given in (35), one can write
 the corresponding symplectic two-form  as
\begin{equation}
\omega_0= \sum_{i=1}^{r} d\xi^{i} \wedge d\xi^{i+r} = \sum_{i=1}^{r} dq^{i} \wedge dp^{i}
\end{equation}
where $r = k+n$ and the real canonical coordinates $\xi^i$ ($i = 1, 2, \cdots , 2r$)
are defined by
\begin{eqnarray}
&& \frac{w^i}{\sqrt{1 + \bar w \cdot w}}=  \frac{1}{\sqrt{2p}}(\xi^{i} + i \xi^{i+r}), \qquad
i = 1, 2, \cdots, n
\nonumber \\
&& \frac{z^i}{\sqrt{1 + \bar z\cdot z}}=  \frac{1}{\sqrt{2q}}(\xi^{i+n} + i \xi^{i+r+n}),
\qquad i = 1, 2, \cdots, k.\nonumber
\end{eqnarray}

Assuming that
%We now assume that
(55) is modified due to the presence of
some external electromagnetic background. More precisely, %This can be formulated by
we replace the canonical two-form $\omega_0$ by a closed new one, such as
\begin{eqnarray}
\omega = \omega_0- \frac{1}{2}\sum_{i,j=1}^{r} {\cal B}_{ij}
d\xi^{i} \wedge d\xi^{j}+ \frac{1}{2}\sum_{i,j=1}^{r}{\cal
E}_{ij} d\xi^{i+r}\wedge d\xi^{j+r}
\end{eqnarray}
where the  antisymmetric tensors ${\cal B}_{ij}$ and ${\cal E}_{ij}$ are phase space
variables dependants. In addition, we  also assume that
the tensor ${\cal B}_{ij}$ and ${\cal E}_{ij}$, encoding the deformation, are
functions of
the phase space coordinates $\xi^{i}$ and $\xi^{i+r}$ $( i = 1, 2, \cdots, r)$, respectively.
These are
\begin{eqnarray}
&& {\cal B}_{ij}\equiv {\cal B}_{ij}(\xi^{1},\xi^{2},\cdots,\xi^{r})
\nonumber\\
&& {\cal E}_{ij}\equiv {\cal
E}_{ij}(\xi^{r+1},\xi^{r+2},\cdots,\xi^{2r})\nonumber.
\end{eqnarray}
The $U(1)$ connections ${\cal B}$ and ${\cal E}$
\begin{eqnarray}
{\cal B} = dA  \qquad {\cal E} = d\bar A
\end{eqnarray}
are given
in terms of  the gauge fields
\begin{eqnarray}
A = \sum_{i=1}^{r}A_i(q)dq^i, \qquad  \bar A = \sum_{i=1}^{r}\bar A_i(p)dp^i
\end{eqnarray}
where the notation bar does not mean the complex conjugate. Alternatively, in compact
form, (56) can be written as
\begin{eqnarray}
\omega = \frac{1}{2}  \omega_{IJ} d\xi^{I}
\wedge d\xi^{J}
\end{eqnarray}
or equivalently, in matrix representation as
\begin{displaymath}
(\omega) =
\left(\begin{array}{ccc}
-{\cal B} & {\bf 1}_{r\times r} \\
-{\bf 1}_{r\times r} & {\cal E} \\
\end{array}\right)
\end{displaymath}
where ${\bf 1}_{r\times r}$ stands for the identity matrix. Two-form
$\omega$ is nondegenerate, i.e. $\det~\omega \neq 0$, when the antisymmetric
tensors ${\cal E}_{ij}$ and ${\cal B}_{ij}$ satisfy the condition
$\det( {\bf 1}_{r\times r} - {\cal E}{\cal B}) \neq 0$. We assume
that such  condition is satisfied. To find the classical equations
of motion and  establish the connection between the classical and
quantum theory, it is necessary to define the Poisson brackets
associated with the new phase space geometry in a consistent way.
Indeed, recalling that the Poisson brackets for the coordinates on
the phase space are the inverse of the symplectic form as matrix. This gives
\begin{eqnarray}
\{{\cal F} , {\cal G}\} = \sum_{I,J} (\omega^{-1})^{IJ}\frac{\partial
{\cal F}}{\partial\xi^{I}}\frac{\partial {\cal G}}{\partial\xi^{J}}
\end{eqnarray}
where $(\omega^{-1})^{IJ}$ is the inverse matrix of $\omega_{IJ}$  and
$({\cal F}, {\cal G})$ are two functions defined on the phase space. Using the matricial form of $\omega$,
it easy to determine the corresponding  inverse. One can see that the
Poisson brackets take the simple form
\begin{eqnarray}
\{{\cal F}, {\cal G}\} = \sum_{ik} (\omega^{-1}_1)_{ik}\frac{\partial
{\cal F}}{\partial \xi^i}\bigg[ \frac{\partial {\cal G}}{\partial
\xi^{k+r}} - \sum_{j} {\cal E}_{kj}\frac{\partial {\cal G}}{\partial
\xi^j} \bigg] -  (\omega^{-1}_2)_{ik}  \frac{\partial {\cal
F}}{\partial \xi^{i+r}}\bigg[ \frac{\partial {\cal G}}{\partial \xi^k}-
\sum_j{\cal B}_{kj}\frac{\partial {\cal G}}{\partial \xi^{j+r}}\bigg]
\end{eqnarray}
where the
matrix elements of the matrices are defined by
\begin{eqnarray}
&& (\omega_1)_{ij} = \delta_{ij} - \sum_{k} {\cal E}_{ik}{\cal B}_{kj} \nonumber \\
&& (\omega_2)_{ij} = \delta_{ij} - \sum_{k}{\cal B}_{ik}{\cal E}_{kj}.
\end{eqnarray}
These  can be read in matrices form as $\omega_1 = {\bf 1}_{r\times r} -
{\cal E}{\cal B}$ and $\omega_2 = {\bf 1}_{r\times r}- {\cal B}{\cal E}$,
respectively. It follows that the modified canonical Poisson
brackets are
\begin{eqnarray}
&& \{ \xi^i , \xi^j\} = - \sum_k(\omega^{-1}_1)_{ik}{\cal
E}_{kj}
\nonumber \\
&& \{ \xi^{i+r} , \xi^{j+r}\} =  \sum_k(\omega^{-1}_2)_{ik}{\cal
B}_{kj}
\\
&& \{ \xi^i , \xi^{j+r}\} =  (\omega^{-1}_1)_{ij} =
(\omega^{-1}_2)_{ji}.\nonumber
\end{eqnarray}
According to the modification of the symplectic structure of the phase
space, we introduce the vector fields $X_{\cal F}$ associated to a
given functional  ${\cal F}(\xi^i , \xi^{i+r})$
\begin{eqnarray}
X_{\cal F} = \sum_{i} X^i \frac{\partial }{\partial \xi^i} +
Y^i\frac{\partial }{\partial \xi^{i+r}}
\end{eqnarray}
such that the interior contraction of $\omega$ with  $X_{\cal F}$
gives
\begin{eqnarray}
{\it i}(X_{\cal F}) \omega = d {\cal F}.
\end{eqnarray}
A straightforward calculation leads
\begin{eqnarray}
 && X^i= \sum_j(\omega^{-1}_1)_{ij}\left(\frac{\partial {\cal
F}}{\partial \xi^{j+r}} - \sum_k{\cal E}_{jk} \frac{\partial {\cal F}
}{\partial \xi^k}\right)
\nonumber \\
 && Y^i=- \sum_j(\omega^{-1}_2)_{ij}\left(\frac{\partial {\cal
F}}{\partial \xi^j} - \sum_k{\cal B}_{jk} \frac{\partial {\cal F}
}{\partial \xi^{k+r}}\right)
\end{eqnarray}
 and one can check
\begin{eqnarray}
{\it i}(X_{\cal F}){\it i}(X_{\cal G}) \omega = \{{\cal F}, {\cal
G}\}.
\end{eqnarray}
As illustration of the above construction, let us  consider the configuration
with
$r=2$, ${\cal E}_{ij}$ and
${\cal B}_{ij}$ are two constants, such as
\begin{eqnarray}
{\cal E}_{ij} = \theta {\epsilon}_{ij}, \qquad {\cal B}_{kj} =
\bar{\theta}{\epsilon}_{ij}
\end{eqnarray}
where $\epsilon_{ij}$ is the usual antisymmetric tensor
($\epsilon_{12} = - \epsilon_{21} = 1$). In this case,
(63) becomes
\begin{eqnarray}
&& \{ \xi^i , \xi^j\} = -
\frac{\theta}{1+\theta\bar{\theta}}\epsilon_{ij}
\nonumber \\
&& \{ \xi^{i+r} , \xi^{j+r}\} =
\frac{\bar{\theta}}{1+\theta\bar{\theta}}\epsilon_{ij}
\\
&& \{ \xi^i , \xi^{j+r}\} = \frac{1}{1+\theta\bar{\theta}}\delta_{ij}\nonumber.
\end{eqnarray}
which are
reflecting a deviation from the canonical brackets. At this stage,
it is remarkable that the modified symplectic form (56) and
corresponding Poisson brackets can be converted in the canonical
forms. This can be achieved by a dressing transformation based on the Moser's lemma [17].
%This is will be developed in the next.

%%%%%%%%%%%%%%%%%%%%%%%%%%%%%%%%%%%%%%%%%%%%%%%%%%%%%%%%%%%%%%%%%%%%%%%%%%%%%%%%%%%%%
\subsection{Symplectic dressing and Moser's lemma}
%%%%%%%%%%%%%%%%%%%%%%%%%%%%%%%%%%%%%%%%%%%%%%%%%%%%%%%%%%%%%%%%%%%%%%%%%%%%%%%%%%%%%

As mentioned above, the Moser's lemma provides us with a nice procedure to locally eliminate
the fluctuations of the symplectic structure. Let us first revisit the derivation
of such lemma in order to give as possible as  general expression of the dressing
transformation, which maps the modified two-form to the original one. Indeed, we consider the most
general case where the matrix elements
$$ (\omega_0)_{IJ} \equiv (\omega_0)_{IJ} (\xi)$$
are nonconstants. We assume that the fluctuation is induced by a closed two-form
$$ F = dA, \qquad F = A_I(\xi)d\xi^I$$
where the summation over repeated indices is understood throughout this subsection.

The Moser's
lemma tells us that  always there exists a diffeomorphism on the phase space $\phi$
whose pullback maps $\omega$ to $\omega_0$
\begin{eqnarray}
\phi^{\ast}(\omega_0+ F) = \omega_0
\end{eqnarray}
which is
\begin{eqnarray}
\phi:  \xi^I \longmapsto \phi(\xi^I), \qquad \frac{\partial\phi(\xi^K)}
{\partial\xi^I}\frac{\partial\phi(\xi^L)}{\partial\xi^J} \omega_{KL}(\phi(\xi))= {\omega_0}_{IJ}(\xi).
\end{eqnarray}
To find out this change of coordinates, one can start by defining a family of one parameter  of symplectic forms
\begin{eqnarray}
\omega(t) = \omega_0+ t F
\end{eqnarray}
interpolating $\omega_0$ and $\omega_0 + F$ for $t=0$ and $t=1$, respectively, with
 $0 \leq t \leq 1$.
Note that, $t$ is just an affine parameter labeling the flow generated by a smooth $t$-dependent vector field $X(t)$.
Accordingly, one  also define a family
of diffeomorphisms
\begin{eqnarray}
\phi^{\ast}(t)\omega(t) = \omega_0
\end{eqnarray}
satisfying $\phi^{\ast}(t=0)= {\rm id}$ and the diffeomorphism $\phi^{\ast}(t=1)$
will be then the required solution of our problem, i.e. (70).
 Differentiating (73), one check that
 $X(t)$  must satisfy the identity
\begin{eqnarray}
0 = \frac{d}{dt} \left[\phi^{\ast}(t)\omega(t)\right] = \phi^{\ast}(t)\left[L_{X(t)}\omega(t) + \frac{d\omega(t)}{dt}\right]
\end{eqnarray}
where $L_{X(t)}$ denotes the Lie derivative of $X(t)$. Using the Cartan identity $L_{X} = ~ \iota_X\circ d + d\circ \iota_X$
and the fact that $d\omega (t) = 0$ to obtain
\begin{eqnarray}
 \phi^{\star}(t)\left\{d\left[\iota_{X(t)}\omega(t)\right] + F\right\}=0
\end{eqnarray}
where $\iota_X$ stands for interior contraction as above. It
follows that  $X(t)$ is satisfying
\begin{eqnarray}
\iota_{X(t)}\omega(t) + A = 0.
\end{eqnarray}
Therefore, the components of  $X(t)$ are given by
\begin{eqnarray}
X^I(t) = - A_J \omega^{-1 JI} (t).
\end{eqnarray}

 For small fluctuations of the symplectic structure, i.e. $F \ll \omega_0$, one can write the inverse of $\omega$ as
\begin{eqnarray}
 \omega^{-1}(t) = \omega_0^{-1} - t\omega_0^{-1} F \omega_0^{-1} + t^2 \omega_0^{-1} F \omega_0^{-1} F \omega_0^{-1} + \cdots.
\end{eqnarray}
This determines the components of  $X(t)$ in terms of the $U(1)$ connection $A$
and its derivatives. It allows us to write down the explicit form of the transformation $\phi$. Indeed,
 since the $t$-evolution of $\omega(t)$ is governed by the first order differential equation
\begin{eqnarray}
 \left[\partial _t + X(t)\right]\omega(t) = 0
\end{eqnarray}
 it is easy to see
\begin{eqnarray}
 \left[\exp(\partial _t + X(t))\exp(-\partial _t )\right]\omega(t+1) = \omega(t).
\end{eqnarray}
Using this, one can  verify the relation
\begin{eqnarray}
 \left[\exp(\partial _t + X(t))\exp(-\partial _t )\right]\vert_{(t=0)}(\omega_0+ F) = \phi^{\ast}( \omega_0+ F ) = \omega_0
\end{eqnarray}
from which one obtains the pull-back mapping $\omega$ to $\omega_0$. This is
\begin{eqnarray}
\phi^{\ast} = {\rm id} + X(0) + \frac{1}{2} (\partial_t X)(0) + \frac{1}{2} X^2(0) + \cdots .
\end{eqnarray}
More explicitly, using  (77) and (78), the contribution of the second term in (82) reads as
\begin{eqnarray}
X(0) = \omega_0^{-1 IJ}A_J\partial_I.
\end{eqnarray}
The contribution of the third term in (82) is given by
\begin{eqnarray}
\frac{1}{2} (\partial_t X)(0) = -\frac{1}{2}(\omega_0^{-1} F \omega_0^{-1})^{IJ}A_J\partial_I.
\end{eqnarray}
Similarly, the last term in (82) leads
\begin{eqnarray}
\frac{1}{2} X^2(0) = \frac{1}{2}(\omega_0^{-1~IJ}A_J\partial_I)(\omega_0^{-1~I'J'}A_{J'}\partial_{I'}).
\end{eqnarray}

Finally, in terms of local coordinates, the coordinate transformation $\phi$
whose pullback maps $\omega_0 + F \longrightarrow \omega_0$ takes the form
\begin{eqnarray}
\phi (\xi^L) = \xi^L + \xi^L_1 + \xi^L_2 + \cdots
\end{eqnarray}
where the second term reads as
\begin{eqnarray}
\xi^L_1 = \omega_0^{-1 LJ}A_J
\end{eqnarray}
and the third one is
\begin{eqnarray}
\xi^L_2 = - \frac{1}{2}\omega_0^{-1 LK} F_{KL'}\omega_0^{-1 L'J}A_{J}
+\frac{1}{2}\omega_0^{- 1IJ}A_J(\partial_I\omega_0^{-1 LJ'})A_{J'} +
\frac{1}{2}\omega_0^{-1 IJ}A_J\omega_0^{-1 LJ'}(\partial_IA_{J'}).
\end{eqnarray}
Using the relations
\begin{eqnarray}
&& \partial_{J'}A_{I'} = (\partial_{J'}\omega_{0I'I})\xi^{I}_1 + \omega_{0I'I}(\partial_{J'}\xi^{I}_1)
\\
&& \partial_I\omega_0^{-1 LJ'} = -\omega_0^{- 1 LJ"}(\partial_I\omega_{0 J"K})\omega_0^{- 1 KJ'}
\end{eqnarray}
and the antisymmetry property of the symplectic form, recall that $\omega_0$
is assumed closed and with non constants matrix elements, one can verify
\begin{eqnarray}
\lefteqn{
\xi^L_2 =-\omega_0^{-1 LK}F_{KL'}\xi^{L'}_1 + \frac{1}{2} \omega_0^{-1 LK}
\omega_0^{-1 MJ}A_J \omega_0^{-1 NJ'}A_{J'}\partial_M\omega_{0 NK}+{}}
\nonumber\\
& & {}+ \frac{1}{2}
\omega_0^{-1 LK} \omega_0^{-1 MS}A_S \omega_{0 MN}
\partial_K(\omega_0^{-1 NS'}A_{S'}).
\end{eqnarray}
It is remarkable that the above dressing transformation  coincides with the
Susskind map derived in connection with  the quantum Hall systems and
noncommutative Chern--Simons theory [30]. Moreover, it leads to  the very familiar
Seiberg--Witten map [18] used in the context of the string and noncommutative
gauge theories. %This will be clarified in the next subsection.

%%%%%%%%%%%%%%%%%%%%%%%%%%%%%%%%%%%%%%%%%%%%%%%%%%%%%%%%%%%%%%
\subsection{{Seiberg--Witten and/or Susskind map }}
%%%%%%%%%%%%%%%%%%%%%%%%%%%%%%%%%%%%%%%%%%%%%%%%%%%%%%%%%%%%%

One can see from  (86-88) that the dressing transformation can
be written as
\begin{eqnarray}
\phi (\xi^L) = \xi^L +  {\hat A}^L
\end{eqnarray}
where the coordinates fluctuations  is the sum of the quantities (87) and (91) given by
\begin{eqnarray}
{\hat A}^L  &=& { \omega_0}^{-1~LK}\left[A_K
- F_{KL'}\omega_0^{-1 L'M}A_M + \frac{1}{2} {\omega_0}^{-1 MJ}A_J {\omega_0}^{-1 NJ'}A_{J'}\partial_M\omega_{0NK}\right]
\nonumber\\
& & + \frac{1}{2} {\omega_0}^{-1~LK}\left[ {\omega_0}^{-1 MS}A_S \omega_{0 MN} \partial_K ({\omega_0}^{-1 NS'}A_{S'})\right].
\end{eqnarray}
The relation (92) can be viewed as a generalization of Susskind map [30]. It encodes the
geometrical fluctuations induced by the external magnetic field $F$.

Moreover,
 (92) coincides also with the Seiberg--Witten map
in a curved manifold  for a noncommutative abelian gauge theory [19].
Indeed, under the gauge transformation
\begin{eqnarray}
A \longrightarrow A + d\Lambda
\end{eqnarray}
 the components (93) transform as
 \begin{eqnarray}
{\hat A}^L \longrightarrow {\hat A}^L +
 \omega_0^{- 1 LJ}\partial_J {\hat \Lambda} + \{ {\hat A}^L , {\hat \Lambda}\}+ \cdots
\end{eqnarray}
where the noncommutative gauge parameter $\hat{\Lambda}$
\begin{eqnarray}
{\hat \Lambda} = \Lambda + \frac{1}{2} \omega_0^{-1 IJ}A_J\partial_I\Lambda + \cdots.
\end{eqnarray}
is expressed
in terms of $\Lambda$ and the abelian connection $A$.
(94), (95) and (96) are the semiclassical version of the Seiberg--Witten map. The connection
$\hat A$ is an induced noncommutative gauge field given in terms
of its commutative counter part $A$.  This establish clearly the above mentioned relation
between Moser's  lemma and Seiberg--Witten map and gives a
correspondence between the symplectic deformations and noncommutative gauge theories.

At this stage, we have the necessary tools to map  two-form (56) to (55). Indeed,
using  (86)  together with  (87) and (91), one obtains the coordinates
change such that $\omega_0 + {\cal E} + {\cal B}$ takes the canonical form
\begin{eqnarray}
\omega_0+ F =   dQ^i \wedge dP^i
\end{eqnarray}
where the new phase space variables $Q^i$ and $P^i$ are given by
\begin{eqnarray}
&& Q^i = \phi^{-1}(q^i) = q^i + \bar A_i(p) - \sum_{j}A_j(q)\left[
{\cal E}_{ij}(p) - \frac{1}{2}\frac{\partial \bar A_j(p)}{\partial
p_i}\right]+ \cdots
\\
 &&
P^i = \phi^{-1}(p^i) =  p^i - A_i(q) + \sum_{j}\bar A_j(p)\left[
{\cal B}_{ij}(q) + \frac{1}{2}\frac{\partial  A_j(q)}{\partial
q_i}\right]+ \cdots.
\end{eqnarray}
It is interesting to note that for  $\bar A_i(p) = 0$, respectively $A_i(q) = 0$,
we recover the  Darboux transformations. On the other hand,
for $r = 2$ when the gauge potentials (57) are defined as
\begin{eqnarray}
A + \bar A = -\frac{1}{2} (\bar \theta \epsilon_{ij}q_idq_j - \theta
\epsilon_{ij}p_idp_j)
\end{eqnarray}
which is
corresponding to constant electromagnetic fields $F$  given by (68),
the dressing transformation  (98-99) gives
\begin{eqnarray}
&& Q^i = (1 + \frac{3}{8} \theta \bar\theta) q^i +
\frac{\theta}{2}\sum_k\epsilon_{ki}p^k
\\
&& P^i = (1 + \frac{3}{8} \theta \bar\theta) p^i +
\frac{\bar\theta}{2}\sum_k\epsilon_{ki}q^k.
\end{eqnarray}
This result will be compared with that will be derived in section 7, see equation (122),
when we
will deal with an exact dressing transformation (Hilbert--Schmidt orthonormalization
procedure mentioned above), which is applicable in some few particular cases.

%%%%%%%%%%%%%%%%%%%%%%%%%%%%%%%%%%%%%%%%%%%%%%%%%%%%%%%%%%%%%%%%%%%%%%%%%%%%%%%%%%%%%%%%%%%%%%%%%%%%%%
\section{Electromagnetic excitations of $A_n$ quantum Hall droplet }
%%%%%%%%%%%%%%%%%%%%%%%%%%%%%%%%%%%%%%%%%%%%%%%%%%%%%%%%%%%%%%%%%%%%%%%%%%%%%%%%%%%%%%%%%%%%%%%%%%%%%

%%%%%%%%%%%%%%%%%%%%%%%%%%%%%%%%%%%%%%%%%%%%%%%%%%%%%%%%%%%%%%%%%%%%%
\subsection{ Hamiltonian and induced electromagnetic interaction}
%%%%%%%%%%%%%%%%%%%%%%%%%%%%%%%%%%%%%%%%%%%%%%%%%%%%%%%%%%%%%%%%%%%%%

In modifying the symplectic structure, the dynamics becomes
described by two-form $\omega_0 + F$. The dressing transformation converts the dynamical system of
$(\omega_0+ F, {\cal V})\Big|_{qp}$ to $(\omega_0, {\cal V}_A)\Big|_{QP}$ where
we use the old symplectic form but  a different Hamiltonian. This latter, can be obtained by simply replacing
the old phase space variables in terms of the new ones. In this respect, inverting  (98) and (99),
one obtains
\begin{eqnarray}
&& q^i = \phi (Q^i) = Q^i - \bar A_i(P) + \sum_{j}A_j(Q)\bigg[
{\cal E}_{ij}(P) - \frac{1}{2}\frac{\partial \bar A_j(P)}{\partial
P_i}\bigg]+ \cdots
\\
&& p^i = \phi (P^i) =  P^i + A_i(Q) - \sum_{j}\bar A_j(P)\bigg[
{\cal B}_{ij}(Q) + \frac{1}{2}\frac{\partial  A_j(Q)}{\partial
Q_i}\bigg]+ \cdots.
\end{eqnarray}
Using this, to obtain  the required Hamiltonian form to the second order in the $A$'s,
which is
 \begin{eqnarray}
{\cal V}_A  &=&  {\cal V}(Q,P) - \sum_i ({\bar A}_i  {\bar u}_i - A_i  u_i )
+\frac{1}{2}\sum_{ij}\left[ {\bar A}_i {\bar A}_j \frac{\partial{\bar u}_i}{\partial Q_j} +
A_iA_j\frac{\partial u_i} {\partial P_j} -2  {\bar A}_iA_j\frac{\partial u_j} {\partial Q_i} \right]
\nonumber\\
& & + \sum_{ij}A_j\left[
{\cal E}_{ij} - \frac{1}{2}\frac{\partial \bar A_j}{\partial
P_i}\right]\bar u_i - \sum_{ij}\bar A_j\left[
{\cal B}_{ij} + \frac{1}{2}\frac{\partial  A_j}{\partial
Q_i}\right] u_i
+{}\cdots
 \end{eqnarray}
where
 \begin{eqnarray}
 u_i = \frac{\partial {\cal V}}{\partial P_i}, \qquad \bar u_i = \frac{\partial {\cal V}}{\partial Q_i}.
\end{eqnarray}
Here again bar does  not stand for complex conjugate.
It is clear that, the dressing transformation eliminates the fluctuations of the symplectic form,
 which become encoded in
the Hamiltonian. In (105), the excitation potential ${\cal V}$ is
\begin{eqnarray}
{\cal V}\equiv {\cal V}(Q,P) = \frac{1}{2}\sum_{i=1}^r [Q_i^2 + P_i^2]
\end{eqnarray}
and the quantities defined by (106) reduces to $u_i = P_i$ and $\bar u_i = Q_i$.

%%%%%%%%%%%%%%%%%%%%%%%%%%%%%%%%%%%%%%%%%%%%%%%%%%%%%%%%%%%%%%%%%%%%%%%%%%%%%%%%%%%%%%
\subsection{Generalized WZW action }
%%%%%%%%%%%%%%%%%%%%%%%%%%%%%%%%%%%%%%%%%%%%%%%%%%%%%%%%%%%%%%%%%%%%%%

The analysis developed in the previous section can be applied to derive the action
describing the electromagnetic
interaction of a higher dimensional  Hall system. Following similar lines as before, the action
is given by
\begin{equation}
S = \int dt{\rm  Tr}\ \left( i\rho_0 U^{\dag} \partial_t U - \rho_0 U^{\dag} V_AU \right)
\end{equation}
where $V_A$ is  the operator associated to the Hamiltonian function given by~(107).
Note that, the kinetic part of the action remains unchanged. Using similar tools as above,
the action can be rewritten as
\begin{equation}
S = S_0 + S_A
\end{equation}
where the unperturbed action $S_0$ is given by~(52) and the electromagnetic coupling contribution
takes the form
\begin{equation}
S_A = - \int dt\  \left[ \rho_0  \star {\cal V}_A -  ({\cal L}\Phi)  \frac{\partial\rho_0}{\partial
r^2}{\cal V}_A\right].
\end{equation}
The above action can also be written as the sum of two terms:  edge and bulk
contributions. Indeed, the Hamiltonian ${\cal V}_A$ can be expressed as
\begin{equation}
{\cal V}_A = {\cal V}^b_A + {\cal V}^e_A
\end{equation}
where the first term is
 \begin{eqnarray}
{\cal V}^b_A  & =&  {\cal V} - \sum_i ({\bar A}_i  {\bar u}_i - A_i  u_i )
+\sum_{ij}\left[ {\bar A}_i u_j \frac{\partial{ A}_j}{\partial Q_i} \right]
\nonumber\\
& & {}+ \sum_{ij}A_j\left[
{\cal E}_{ij} - \frac{1}{2}\frac{\partial \bar A_j}{\partial
P_i}\right]\bar u_i - \sum_{ij}\bar A_j\left[
{\cal B}_{ij} + \frac{1}{2}\frac{\partial  A_j}{\partial
Q_i}\right] u_i
+{}\cdots
 \end{eqnarray}
and the second reads as
 \begin{eqnarray}
{\cal V}^e_A  =
\frac{1}{2}\sum_{ij}\left[\frac{\partial ({\bar A}_i {\bar A}_j {\bar u}_i)}{\partial Q_i}  + \frac{\partial (A_iA_j u_i)}{\partial P_j}  -2 \frac{\partial ({\bar A}_iA_j u_j)}{\partial Q_i}  \right].
\end{eqnarray}

It follows that the action  (110) can be separated  into two parts
\begin{equation}
S_A = S_A^{\rm bulk} + S_A^{\rm edge}
\end{equation}
 where the edge contribution is  given by
\begin{equation}
S_A^{\rm edge} = - \int dt\  \left[ \rho_0  \star {\cal V}^e_A -  ({\cal L}\Phi)  \frac{\partial\rho_0}{\partial
r^2}{\cal V}_A\right]
\end{equation}
and the bulk term is
\begin{equation}
S_A^{\rm bulk} = - \int dt\   \rho_0  \star {\cal V}^b_A.
\end{equation}
Note that, the quantity (113) involves only derivatives. Then, by expanding the star product
in the first term in (115) and  integrating, one can see that it produces
a boundary contribution. The second  term in (115) is obviously  a boundary contribution because
it involves the derivative of the density, i.e. the derivative of the density
gives a delta function defined on the edge. Clearly, in the absence of electromagnetic
field $ A = 0$, the action (109) reduces to $S_0$ as it should be.

To close this section two remarks are in order. First, the electromagnetic excitation of
${\bf CP}^k$ quantum Hall droplet, with $A_n$ spin degrees of freedom, is generated here
from a purely symplectic point of view. This gives an alternative description to the results
presented in [11] where a similar issue was investigated. Secondly, our model can
easily be adapted to some geometries and situations with other kinds of internal degrees
of freedom.

%%%%%%%%%%%%%%%%%%%%%%%%%%%%%%%%%%%%%%%%%%%%%%%%%%%%%%%%%%%%%%%%%%%%%%%%%%%%%%%%%%%%%%%%%%%%%%%%%%%%%%%%%%%%%%%%%
\section{$A_1$ quantum Hall droplet in sphere} %:  Constant symplectic fluctuation }
%%%%%%%%%%%%%%%%%%%%%%%%%%%%%%%%%%%%%%%%%%%%%%%%%%%%%%%%%%%%%%%%%%%%%%%%%%%%%%%%%%%%%%%%%%%%%%%%%%%%%%%%%%%%%%%%

%%%%%%%%%%%%%%%%%%%%%%%%%%%%%%%%%%%%%%%%%%%%%%%%%%%
\subsection{ Excitation potential}
%%%%%%%%%%%%%%%%%%%%%%%%%%%%%%%%%%%%%%%%%%%%%%%%%%

In this section, we consider the most simplest case where the particles are living
in the sphere ${\bf S}^2$ and the internal degrees of freedom are described by $A_1$
spin algebra. Here the total phase space ${\cal M}$ is four-dimensional space.
As far as the electromagnetic excitation (or symplectic modification) is concerned,
we will restrict ourself
 to two-form given by (68). Instead of the dressing transformation based on
 the Moser's lemma, we make use of another method, which gives in this particular
 case an exact coordinates change.  This is based on the so-called Hilbert--Schmidt
 orthonormalization procedure.
 In the present case,  we rewrite the Poisson brackets (69)  as
\begin{eqnarray}
&& \{ q^i , q^j\} = - \frac{\theta}{1+\theta\bar{\theta}}\epsilon_{ij}
\nonumber\\
&& \{ p^i , p^j\} = \frac{\bar{\theta}}{1+\theta\bar{\theta
}}\epsilon_{ij}
\\
&& \{ q^i , p^j\} = \frac{1}{1+\theta\bar{\theta }}\delta_{ij}\nonumber
\end{eqnarray}
where $ i, j = 1, 2$, they reflect a deviation from the canonical brackets. In addition,
they traduce the noncommutativity
between the coordinates and momenta of the internal and kinematical phase spaces.

The excitation potential (45) coincides for $r=2$ with
 a bidimensional harmonic oscillator. It is simply verified that the deformed symplectic
 two-form $\omega_0 + {\cal B}+ {\cal E}$ can be  converted in a canonical
one by using an analog of the Hilbert--Schmidt transformation
\begin{eqnarray}
&& Q^i = a q^i + \frac{1}{2} b\theta\sum_k\epsilon_{ki} p^k
\nonumber\\
&& P^i = c p^i + \frac{1}{2} d\bar{\theta}\sum_k\epsilon_{ki} q^k
\end{eqnarray}
where
\begin{eqnarray}
a = c = \frac{1}{b} = \frac{1}{d}=
\frac{1}{\sqrt{2}}\sqrt{1+\sqrt{1+\theta\bar{\theta}}}.
\end{eqnarray}
 We assume that  $1 + \theta \bar
\theta  > 0$. Consequently,
the Poisson brackets (117) give the canonical ones
\begin{eqnarray}
&& \{ Q^i , Q^j\} =  0 \nonumber\\
&&  \{ P^i , P^j\} = 0 \\
&& \{ Q^i , P^j\} = \delta_{ij}\nonumber
\end{eqnarray}
and the symplectic two-form  becomes
\begin{eqnarray}
\omega = \sum_{i} dQ^i \wedge dP^i.
\end{eqnarray}
It is interesting to notice that, for small values of $\theta$ and $\bar \theta$,
 (118) and (119) give
\begin{eqnarray}
&& Q^i = (1 + \frac{1}{8} \theta \bar\theta) q^i +
\frac{\theta}{2}\sum_k\epsilon_{ki}p^k
\nonumber\\
&& P^i = (1 + \frac{1}{8} \theta \bar\theta) p^i +
\frac{\bar\theta}{2}\sum_k\epsilon_{ki}q^k
\end{eqnarray}
which are sensitively similar to (101) and (102) those obtained by
using Moser's lemma.  Inverting the transformation (118), we have
\begin{eqnarray}
&& q^i = \frac{a}{\sqrt{1+\theta{\bar\theta}}}\left[ Q^i +
\frac{\theta}{2a^2} \sum_k\epsilon_{ik} P^k\right]
\nonumber\\
&& p^i = \frac{a}{\sqrt{1+\theta{\bar\theta}}}\left[P^i +
\frac{\bar{\theta}}{2a^2}\sum_k\epsilon_{ik} Q^k\right].
\end{eqnarray}

The excitation potential (45) for $ r = 2$ becomes
\begin{eqnarray}
{\cal V} = \frac{a^2}{2(1+\theta{\bar\theta)}}\left[ \sum_{i} (1 +
\frac{\theta^2}{4a^4}) P^iP^i + (1 +
\frac{\bar{\theta}^2}{4a^4})Q^iQ^i + (\frac{\theta}{a^2} -
\frac{\bar{\theta}}{a^2}) \sum_j\epsilon_{ij}Q^iP^j\right].
\end{eqnarray}
Evidently the transformation (118) also eliminates the modification of the symplectic
structure, which become incorporated in the excitation potential. For our purpose,
 the expression (124)
can be cast in a more appropriate form  by introducing the variables
\begin{eqnarray}
z_{\pm} = \frac {\sqrt{\Delta}}{2} \left[(Q^1 \pm i Q^2) +  \frac{i}{\Delta}  (P^1 \pm i P^2)\right]
\end{eqnarray}
where
\begin{eqnarray}
\Delta = \sqrt{\frac{4a^4 +\bar{\theta}^2 }{4a^4 + \theta^2}}.
\end{eqnarray}
Indeed, substituting (125) in (124), we obtain
\begin{eqnarray}
{\cal V} =  \omega_+ z_+\bar z_+ + \omega_- z_-\bar
z_-
\end{eqnarray}
where
\begin{eqnarray}
\omega_{\pm} = \frac{\sqrt{(4a^4+\theta^2)(4a^4 + \bar{\theta}^2
)}}{4a^2(1+\theta\bar \theta)}
~ \mp ~
 \frac{\theta - \bar \theta} {2(1+ \theta \bar \theta )}.
\end{eqnarray}
Note that,  two-form (121) rewrites as
\begin{eqnarray}
\omega = i (dz_+ \wedge d\bar z_+ + dz_- \wedge d\bar z_-).
\end{eqnarray}
Upon quantization, all canonical variables become Heisenberg operators satisfying commutation rules
 according to the canonical prescription: (Poisson bracket $\longrightarrow$ -i commutator). It follows that the
nonvanishing commutators  are
\begin{eqnarray}
[ z_+ , \bar z_+] = 1, \qquad [ z_- , \bar z_-] = 1.
\end{eqnarray}
Note that, the excitation potential is a superposition of two one-dimensional harmonic oscillators.
This  is very important and will have interesting consequences on the electromagnetic excitations
of the quantum Hall effect in four-dimensional phase space.
%This is will be clarified
%in the next section.

%%%%%%%%%%%%%%%%%%%%%%%%%%%%%%%%%%%%%%%%%%%%%%%%%%%%%%%%%%%%%%
\subsection{Electromagnetic excitations of quantum Hall droplets}
%%%%%%%%%%%%%%%%%%%%%%%%%%%%%%%%%%%%%%%%%%%%%%%%%%%%%%%%%%%%

The excitation  potential (127) determines the spacial shape of the density function
\begin{equation}
\rho_0(\bar z_{\pm}, z_{\pm} ) = \Theta( M - (\omega_+ z_+\bar z_+ + \omega_- z_-\bar z_-))
\end{equation}
where $M$ is the total number of particles in the LLL. Following the same
strategy as above, one can evaluate the action governing the edge excitations in
the presence of an electromagnetic background.
The kinetic term in (108) gives
\begin{equation}
 i \int dt {\rm Tr} \left(\rho_0
U^{\dag}\partial_tU\right) \simeq \frac {1}{2} \int
\{\Phi,\rho_0\}\partial_t\Phi
\end{equation}
where the symbol $\{ , \}$ is the
Poisson bracket. It is calculated as
\begin{equation}
\{\Phi , \rho_0\} = (\omega_+{\cal L}_+\Phi + \omega_-{\cal L}_-\Phi)  \frac{\partial\rho_0}{\partial r^2}
\end{equation}
where $r^2 = \omega_+ z_+\bar z_+ + \omega_- z_-\bar z_-$ and
 the first order differential operators are defined by
\begin{equation}
 {\cal L}_{\alpha} =  i \left(z_{\alpha} \frac{\partial}{\partial z_{\alpha}}
- \bar z_{\alpha}\frac{\partial}{\partial \bar z_{\alpha}}\right), \qquad \alpha = + , -.
\end{equation}
The
 derivative of the density function, in (133), gives a  $\delta$ function with support on
the boundary $\partial {\cal D}$ of the droplet ${\cal D}$ defined
by $ r^2 = M$. Then, we have
\begin{equation}
 i \int dt {\rm Tr} \left(\rho_0
U^{\dag}\partial_tU\right) \approx -\frac{1}{2} \int_{\partial {\cal
D}\times{\bf R}^+} dt \left(\omega_+{\cal L}_+\Phi + \omega_-{\cal L}_-\Phi \right) \partial_t\Phi.
\end{equation}
The evaluation of the potential term in (108) gives
\begin{equation}
  \int dt {\rm Tr} \left(\rho_0
U^{\dag}V U\right) \approx \frac{1}{2} \int_{\partial {\cal
D}\times{\bf R}^+} dt \left(\omega_+{\cal L}_+\Phi + \omega_-{\cal L}_-\Phi\right)^2
\end{equation}

Combining (132) and (136), we get
\begin{equation}
S \approx -\frac{1}{2}\int_{\partial {\cal D}\times {\bf R }^+} dt
\left[{\omega}_+({\cal L}_+\Phi)
+ {\omega}_-({\cal L}_-\Phi)\right] \left[( \partial_t\Phi )+{\omega}_+({\cal L}_+\Phi)
+ {\omega}_-({\cal L}_-\Phi)\right].
\end{equation}
This action involves only the time derivative of $\Phi$ and
tangential derivatives $({\cal L}_{\alpha}\Phi)$. It is a generalization
of a chiral abelian Wess--Zumino--Witten (WZW) theory. For $\theta =
0 $ and $\bar\theta = 0$, we get
\begin{equation}
S \approx -\frac{1}{2}\int_{\partial {\cal D}\times {\bf R }^+} dt
\bigg((\partial_t\Phi ) ({\cal L}\Phi) +\omega ({\cal L}\Phi)^2
\bigg).
\end{equation}
where ${\cal L} = {\cal L}_+ + {\cal L}_-$. We
recover then the WZW usual action for the edge states associated with
un-gauged Hall droplets in four-dimensional phase space, see (52).
 It is interesting to stress that the  action (138) is similar
to one derived in [2] for the  quantum Hall droplets on ${\bf CP}^2$.

%%%%%%%%%%%%%%%%%%%%%%%%%%%%%%%%%%%
\subsection{ Edge fields}
%%%%%%%%%%%%%%%%%%%%%%%%%%%%%%%%%%%%%

The action (137) is minimized by the field $\Phi$
satisfying
the equation of motion
\begin{equation}
\sum_{\alpha = \pm}(\omega_{\alpha}{\cal L}_{\alpha})
\left[\partial_t \Phi +\omega_{\alpha}{\cal L}_{\alpha} \Phi\right] = 0.
\end{equation}
The edge field $\Phi$ can be expanded in powers of the phase space variables $z_{\alpha}$.
Note that, since the excitations are moving on the real 3-sphere ${\bf S}^3 \sim SU(2)$,
it is more appropriate to use the $SU(2)$ parameterization. This is
\begin{equation}
\sqrt{\omega_+} ~ z_+ = \sqrt{M}\frac{\sqrt{\bar \zeta \zeta}}{\sqrt{1+\bar \zeta \zeta}}e^{i\phi_+},
\qquad \sqrt{\omega_-}~ z_- = \sqrt{M}\frac{1}{\sqrt{1+\bar \zeta \zeta}}e^{i\phi_-}
\end{equation}
where $\zeta$ and $\bar \zeta$ are the local complex coordinates for $SU(2)$.
The operators ${\cal L}_{\pm}$ reduce to partial
derivatives $\partial_{\phi_{\pm}}$ with respect to $\phi_{\pm}$. Thus,
the field $\Phi$ can be expressed as
\begin{equation}
\Phi = \sum_{n_+,n_-}  c_{n_+,n_-}(t) e^{i\phi_+n_+}e^{i\phi_-n_-}
\end{equation}
where the coefficients $c_{n_+,n_-}$ are $(\phi_+, \phi_-)$-independents for $(n_+ , n_- )\neq (0 , 0)$.
It follows that the general
solution of the equation of motion (139) takes the form
\begin{equation}
\Phi = (\phi_+ - \omega_+ t)+(\phi_- - \omega_- t)+ \sum_{n_+ n_-}  c_{n_+,n_-}(0) e^{i(\phi_+ - \omega_+ t)n_+}e^{i(\phi_- - \omega_- t)n_-}.
\end{equation}
It is clear from the last equation that the noncommutativity arising from the symplectic
modification changes the propagation velocities of the edge field along the angular
directions. It is also important to stress that the velocities $\omega_+$ and $\omega_-$
are different (respectively equal) for $\theta \neq \bar\theta$ (respectively $\theta = \bar\theta$).
Finally note that, for $\theta = \bar \theta = 0$, i.e. $\omega_{\pm} = 1$,
 the field (142)  coincides with the edge excitations of
the ${\bf CP}^2$ quantum Hall droplet derived in [2].

%%%%%%%%%%%%%%%%%%%%%%%%%%%%%%%%%%%%%%%%
\section{Concluding remarks}
%%%%%%%%%%%%%%%%%%%%%%%%%%%%%%%%%%%%%%%%%

We introduced
 the internal phase space
of $A_n$ spin systems starting from the analytical representation of $A_n$ spin states.
Indeed, the finite dimensional vector spaces, on which the spin matrices act,
are provided with an over complete basis labeled by coordinates on the complex projective
space ${\bf CP}^n$. As far as the quantum Hall effect on ${\bf CP}^k$
is concerned, we investigated
the classical phase space structure of $A_n$ particles in the lowest Landau levels.
The total phase space incorporates the internal as well as the kinematical degrees of freedom
of the
Hall system.

For a large collection of $A_n$ particles, forming a
dense Hall droplet, we  evaluated
 the action governing the edge excitations. We showed that  no trivial
 dynamics arose and the action is formally
similar to that obtained for the spinless case. This agrees with the analysis presented
in [13] and is mainly due to the fact that we ignored a possible coupling
between the internal and kinematical phase space variables. Indeed, we described
$A_n$ spin degrees in terms of an internal compact phase space whose canonical structure
is decoupled from the kinematical phase space. In this description, it is clear that any
nonexpected spin dynamics can arise out of the Hamiltonian. In this respect, we
introduced a nonzero phase structure coupling two spaces. This induces a
noncommutative structure in the phase space and can be interpreted as
the electromagnetic excitation of the Hall system.

Using a coordinates change based on the Moser's lemma, we obtained an
elegant procedure to eliminate the fluctuations of the symplectic two-form. In this way,
the effects of the symplectic modifications become incorporated in the Hamiltonian of the system.
We derived the bulk and edge actions describing the electromagnetic excitations of dense configuration
of $A_n$ particles. It is important to stress that our result agrees with that obtained in [11]
for spinless particles. The result presented here deals essentially with symplectic fluctuations.

As illustration, we gave a detailed study for particles, obeying $A_1$ statistics,
living in two-sphere
${\bf S}^2$. For a specific form of the symplectic fluctuation, we obtained an explicit
expression of the edge field excitation. It is clearly shown that the electromagnetic
excitations induces an asymmetry between the two angular directions of the droplet.
Indeed, the edge field propagates with different velocities along these directions.
Finally, we established am interesting relation between the dressing transformation arising from
the Moser's lemma and the Seiberg--Witten map.

%%%%%%%%%%%%%%%%%%%%%%%%%%%%%%%%%%%%%%%%%%%%%%
\section*{Acknowledgments}
%%%%%%%%%%%%%%%%%%%%%%%%%%%%%%%%%%%%%%%%%%%%

MD would like to express his thanks to
Max Planck Institute for Physics of Complex Systems (Dresden-Germany) where this work was done.
AJ is thankful to Dr. Abdullah Aljaaferi for his help and support.

\end{document}